\documentclass[final,3p,times,twocolumn]{elsarticle}

\usepackage{amssymb}
\usepackage{amsmath}

\usepackage[english]{babel}

\usepackage{amsmath}
\usepackage{graphicx}
\usepackage[colorlinks=true, allcolors=blue]{hyperref}
\usepackage{multicol}

\journal{Computational Materials Science}

\begin{document}

\begin{frontmatter}


\title{Symmetry-aware Conditional Generation of Crystal Structures Using Diffusion Models}

\author{Takanori Ishii\corref{cor2}\fnref{label1}}
\author{Kaoru Hisama\corref{cor1}\fnref{label1}}
\ead{hisama@preferred.jp}
\author{Kohei Shinohara\fnref{label1}}


\cortext[cor2]{Internship student at Preferred Networks Inc.}
\cortext[cor1]{Corresponding Author}
\affiliation[label1]{organization={Preferred Networks Inc.},
            addressline={1-6-1, Otemachi, Chiyoda-ku}, 
            city={Tokyo},
            postcode={100-0004}, 
            country={Japan}}





\begin{abstract}

The application of generative models in crystal structure prediction (CSP) has gained significant attention. Conditional generation—particularly the generation of crystal structures with specified stability or other physical properties has been actively researched for material discovery purposes. Meanwhile, the generative models capable of symmetry-aware generation are also under active development, because space group symmetry has a strong relationship with the physical properties of materials. In this study, we demonstrate that the symmetry control in the previous conditional crystal generation model may not be sufficiently effective when space group constraints are applied as a condition. To address this problem, we propose the WyckoffDiff-Adaptor, which embeds conditional generation within a WyckoffDiff architecture that effectively diffuses Wyckoff positions to achieve precise symmetry control. We successfully generated formation energy phase diagrams while specifying stable structures of particular combination of elements, such as Li–O and Ti–O systems, while simultaneously preserving the symmetry of the input conditions. The proposed method with symmetry-aware conditional generation demonstrates promising results as an effective approach to achieving the discovery of novel materials with targeted physical properties.

\end{abstract}

\begin{graphicalabstract}
\centering
\includegraphics[width=1.5\linewidth]{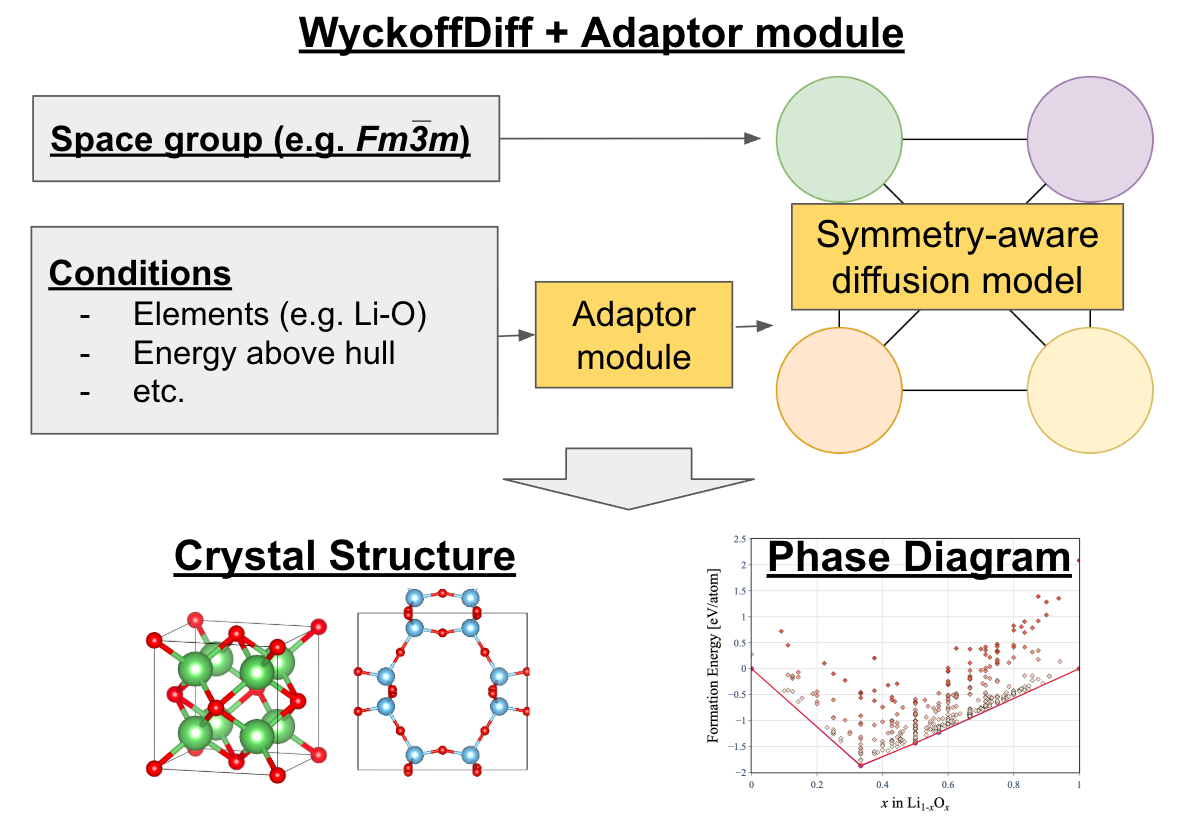}
\end{graphicalabstract}

\begin{highlights}

\item Conditional generation of crystal structures using symmetry-aware Wyckoff position-based architecture is realized.
\item The proposed model improved the control of symmetry even after structure optimization.
\item The zero-energy phase diagram task with given elements (Li–O) was successful.
\item Some limitations of the method were also revealed.

\end{highlights}


\end{frontmatter}




\section{Introduction}

In materials discovery, crystal structure prediction (CSP), which is a computational method for finding new crystal structures, represents a fundamental and critical task that supports both upstream synthetic planning and experimental synthesis \cite{Oganov2019,Oganov2010_textbook}. When searching for experimentally unknown crystals, computational approaches such as CSP using theoretical calculations e.g., first-principles calculations, or empirical potentials have become increasingly important. In addition, methods for an autonomous exploration of new materials with a combination of theoretical prediction and laboratory automation have recently been reported remarkably \cite{Merchant2023,Szymanski2023}.
While CSP plays a vital role in predicting stable crystal structures, considering various polymorphic forms and compositional variations expands the search space exponentially, making computational methods essential for efficient exploration. Algorithms for generating novel crystal structures typically employ metaheuristic approaches, including random structure search, genetic algorithms, and Bayesian optimization, and many software packages have been developed for integrating these methods, such as USPEX \cite{LYAKHOV20131172}, XtalOpt \cite{LONIE2011372}, CALYPSO \cite{WANG20152580}, CrySPY \cite{Yamashita2021} and others as one of the authors proposed recently \cite{shibayama2025efficientcrystalstructureprediction}. However, there are still difficulties to overcome to search for novel and stable crystal structures from the vast degree of freedom of atom compositions, positions and shape of crystal lattice cells, and their symmetry with 230 space-groups.

 Recent years have seen growing interest in alternative approaches using generative machine-learning models to predict candidate crystal structures, such as variational autoencoders (VAEs) and diffusion models such as CDVAE \cite{xie2022}, or DiffCSP \cite{jiao2024crystalstructurepredictionjoint}. These methods aim to navigate the vast configurational spaces more efficiently by generating promising candidate structures. Those approaches are still under development \cite{handoko2025artificialintelligencegenerativemodels}, and benchmarking studies started to evaluate whether the models truly generate novel structures \cite{negishi2025continuous, duval2025lematgenbench}. Among these, MatterGen \cite{MatterGen2025} was particularly notable as a model capable of conditional generation, which enables predictions of structures with desired properties provided with input labels (e.g., composition, energy as stability, band gap). 

 One key limitation of conventional approaches such as CDVAE or MatterGen is their inadequate treatment of crystal symmetry, particularly space-group symmetries. Because space groups are strongly linked to materials properties through structure–property relationships \cite{Nye1985-ax,Suzuki2022,de2012structure, callister2020materials}, symmetry-aware generative models have been developed in recent years, including the diffusion-based WyckoffDiff \cite{kelvinius2025wyckoffdiffgenerativediffusion} and the flow-matching-based Space-Group Conditional Flow Matching (SGFM) \cite{puny2025spacegroupconditionalflow}. These models incorporate symmetry by utilizing Wyckoff positions—represented using fractional coordinates according to space group symmetry.

This study aims to clarify the limitations of conditional generation methods for crystal structure generation that incorporate space group symmetry, while simultaneously establishing a multiobjective conditional generation approach that integrates symmetry-aware modeling using the Wyckoff positions. First, we evaluated the performance of a MatterGen-based model trained with space group constraints, finding that the built-in adapter module alone was insufficient for controlling symmetry. We then developed a WyckoffDiff-based model featuring an adapter module similar to MatterGen. This model is capable of both energy-condition-based conditional generation and producing structures with specified space groups. The proposed WyckoffDiff-Adapter demonstrated the ability to preserve the symmetry specified during input. Furthermore, the model successfully reproduced the formation energy phase diagrams for binary systems (Li–O, Ti–O) with significantly faster computational times compared to the MatterGen-based model. The WyckoffDiff-Adapter approach presents a viable solution for achieving symmetry-aware conditional generation of novel crystals with desired symmetry, energy stability, and other physical properties. However, limitations were also identified, including the possibility of generating chemically unsound bond configurations and the inherent limitation that Wyckoff position prediction becomes arbitrary in the WyckoffDiff-based architecture.

\section{Computational Methods}

\subsection{WyckoffDiff-Adaptor architecture}

To realize conditional generation of crystals under a symmetry-aware architecture, we implemented adaptor modules into the WyckoffDiff model. The fundamental concept of WyckoffDiff \cite{kelvinius2025wyckoffdiffgenerativediffusion} involves first determining crystal symmetry before arranging atoms, which addresses the conventional crystal generation model's tendency to produce structures with low symmetry space groups (specifically P1). By explicitly incorporating symmetry into the input, WyckoffDiff employs a discrete diffusion model that directly processes the crystal's symmetry information, including both space group and Wyckoff positions. Since it explicitly uses symmetry as input and generates protostructures, i.e., element-by-Wyckoff position occupancy information, WyckoffDiff differs from generation models like MatterGen that produce coordinate-based structures in that it guarantees the symmetry of generated crystal structures.

In WyckoffDiff, crystal structures are defined as a protostructure, 
$  \boldsymbol{M} = (s, \boldsymbol{z}^\infty, \boldsymbol{z}^0) $.
Here, $s$ represents the space group, $\boldsymbol{z}^\infty$ denotes unconstrained  Wyckoff positions (general Wyckoff positions in International Tables for Crystallography \cite{ITA}) with degrees of freedom, and $\boldsymbol{z}^0$ represents restricted Wyckoff positions (special Wyckoff positions in International Tables for Crystallography).
The restricted Wyckoff positions in this context refer to positions where some constraints are imposed on all coordinate axes $(x, y, z)$, while all other Wyckoff positions without such constraints are defined as general Wyckoff positions. WyckoffDiff trains its model by performing discrete diffusion inverse generation in the following manner. (1) For general Wyckoff positions, it generates a matrix $\boldsymbol{z}^\infty$ representing atom species $\times$ count. (2) For restricted Wyckoff positions, it generates a vector $\boldsymbol{z}^0$ representing atom species + 0 (vacant sites)
(3) After (1) and (2), it uses these outputs to generate crystal structures through discrete diffusion.

\begin{figure*}
    
\centering
\includegraphics[width=0.9\linewidth]{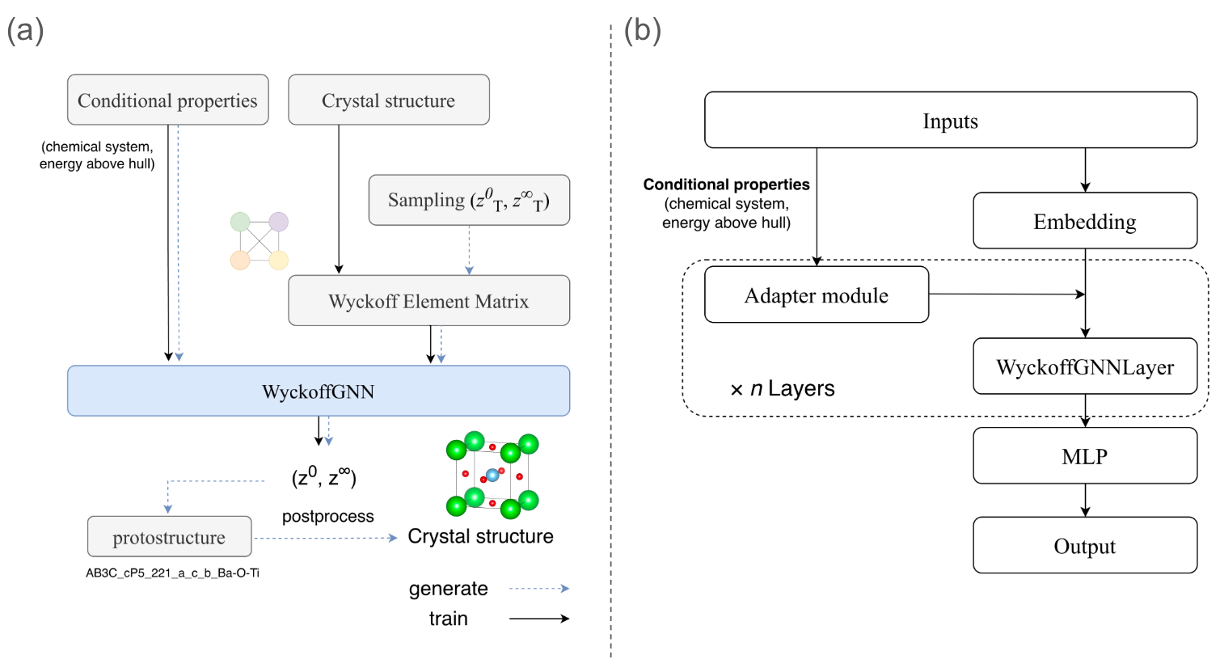}
\caption{(a) Schematic diagram of the training and generation process of the WyckoffDiff-Adaptor, and (b) architecture of WyckoffGNN model including the adaptor module.}
\label{fig:arch}
\end{figure*}

The original WyckoffDiff implementation does not support conditional generation using material property constraints. To enable this functionality through MatterGen's Adapter module, we implemented an Adapter module within WyckoffDiff itself. The WyckoffDiff-Adaptor is based on the MatterGen codebase. Figure \ref{fig:arch} (a) shows the flow of the training and generation process. The training crystal structures are converted to Wyckoff element matrices similar to the original WyckoffDiff, which denote the occupation of the element within the protostructure labels. The conditional properties such as chemical systems, and the energy above hull are also embedded to the WyckoffGNN layers, the main neural network of this model. The WyckoffGNN layer outputs the Wyckoff element matrix. In the generation post-process, it is converted to a protostructure label \cite{Mehl2017,Hicks2019,Hicks2021,Eckert2024}, which is available in Pymatgen \cite{Ong2013} or Aflow \cite{Calderon2015} packages and converted to the conventional crystal structure with atom positions and the cell information using PyXtal. Figure \ref{fig:arch} (b) shows the complete architecture diagram of the improved WyckoffDiff with integrated Adapter module. The conditional properties are embedded by the Adaptor modules during the $n$ layers of the original WyckoffGNN layer connections.

The diffusion model architecture was configured with the following parameters. The maximum diffusion time step $T$ was set to 1000. For the compositional input, the maximum number of chemical elements is set to 100 ($\text{num\_elements} = 100$), with a maximum number of atoms per element set to 54 ($\text{max\_num\_atoms} = 54$). The core denoising network was  configured with 3 GNN layers ($L=3$). The hidden dimension of the atomic feature vectors $h^l_i$ was set to 256 ($\text{hidden\_dim} = 256$). The dimensionality of the position-specific embedding $h^{\text{pos}}_i$ was 16 ($\text{dof\_pos\_sg\_emb\_size} = 16$). The SiLU (Sigmoid Linear Unit) was employed as the activation function throughout the network. Finally, the Adaptor module, responsible for processing compositional data, utilized a hidden layer dimension of 64 ($\text{adapter\_prop\_dim} = 64$).

\subsection{Dataset, training and validation setup}

The numerical experiment is performed by training both MatterGen and WyckoffDiff, specifying three conditions each: space group, energy above the hull, and atomic species. The training dataset utilizes the MP-20 dataset \cite{Jain2013, xie2022}, which was also employed in both MatterGen and WyckoffDiff. For the proposed WyckoffDiff-Adaptor, the learning rate of the base model and the finetuning model with the Adaptor module are set to $\mathrm{10^{-4}}$, and $\mathrm{2\times10^{-4}}$, respectively. The batch size of the training crystal structures are set to $32$, and the epochs of the base model and the fine tuning model are $900$, and $200$, respectively.

The conditional fine-tuned models are tested for the tasks to explore the stable structures in formation energy phase diagrams of Li–O and Ti–O systems. The generated structures are optimized using MatterSim \cite{yang2024mattersim} to assess structural stability. The phase diagram tasks are executed under the input conditions are stable structure (energy above hull = 0), the chemical system (Li–O, and Ti–O) and the input set of space groups are specified based on the 20 most frequent space groups in the mp-20 dataset. Generation conditioned by larger numbers of elements is also performed for ternary Al-Mg-O, and  quaternary Ba-Ta-In-O systems. The maximum number of post-processing attempts of the WyckoffDiff-Adaptor using PyXtal was set to 100, and if the  protostructure of the generated crystal structure did not match the input protostructure, the structure generation was rejected.

The VESTA software \cite{Momma2011} is used for visualization of the crystal structure in Figure \ref{fig:arch}, Figure \ref{fig:structures_binary}, and \ref{fig:structures}. The Pymatgen phase diagram module \cite{Ong2008,Ong2010} is used for the visualization of the phase diagrams in Figure \ref{fig:Li-O}.

\section{Results and Discussion}

\begin{figure*}
\centering
\includegraphics[width=0.65\linewidth]{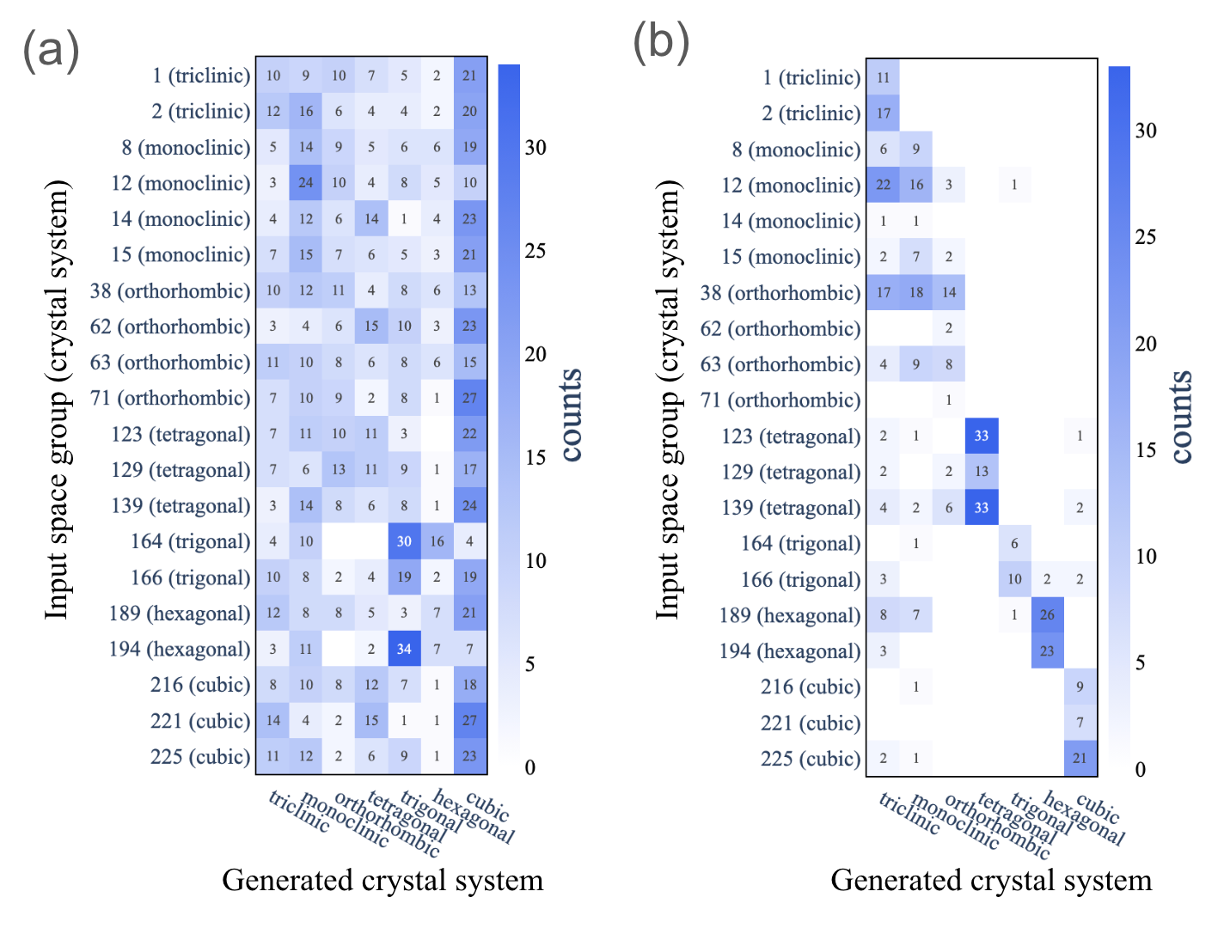}
\caption{Heatmaps of the numbers of generated structure between input and generated structures, using (a) MatterGen and (b) WyckoffDiff-Adaptor, both conditioned by space group. The rows denote the space group of the input and the columns denote the crystal system of the generated structures after optimization using MatterSim. Note that the number of generated structure by WyckoffDiff-Adaptor for each input space group is less than the generation batch size $64$, because the generated protostructure is duplicated.}
\label{fig:spacegroup}
\end{figure*}

The stable structures are explored for the Li–O chemical system, generating total 1280 structures for each model (Mattergen/WyckoffDiff-Adaptor). Figure \ref{fig:spacegroup} (a) displays the heatmap of the crystal system corresponding to the specified space group from MatterGen conditioning generation and the subsequent structural optimization using MatterSim. The vertical axis represents the space group specified during conditioning generation, while the horizontal axis indicates the resulting crystal system after structural optimization using MatterSim. Structures where the diagonal elements of the heatmap exhibit darker colors indicate better results, as they maintain the specified space group classification even after structural optimization.
The heatmap shows that while structures in the trigonal and monoclinic crystal systems somewhat preserve the specified space group classification after optimization, a macro-level overview reveals that either the specified space group was never generated in the first place, or the structural optimization process resulted in structures that fall under crystal systems that cannot be classified using the specified space group. This suggests that controlling symmetry when generating crystal structures using MatterGen proves challenging.
On the other hand, Figure \ref{fig:spacegroup} (b) shows the result using the proposed model of the WyckoffDiff-Adaptor. The diagonal components show darker colors, indicating that even when crystal structures are generated with specified space groups through conditional generation followed by structural optimization, they maintain the crystal system to which their specified space group belongs.
This demonstrates that WyckoffDiff achieves better preservation of the specified symmetry than MatterGen in conditional generation.

The results suggest that MatterGen faces significant challenges in producing crystal structures that maintain symmetry if the target space group is given as input. To improve symmetry control capabilities, the symmetry-aware architecture of the WyckoffDiff, which explicitly incorporates symmetry information such as Wyckoff positions as input parameters, yielded better results.

\begin{figure*}
\centering
\includegraphics[width=0.75\linewidth]{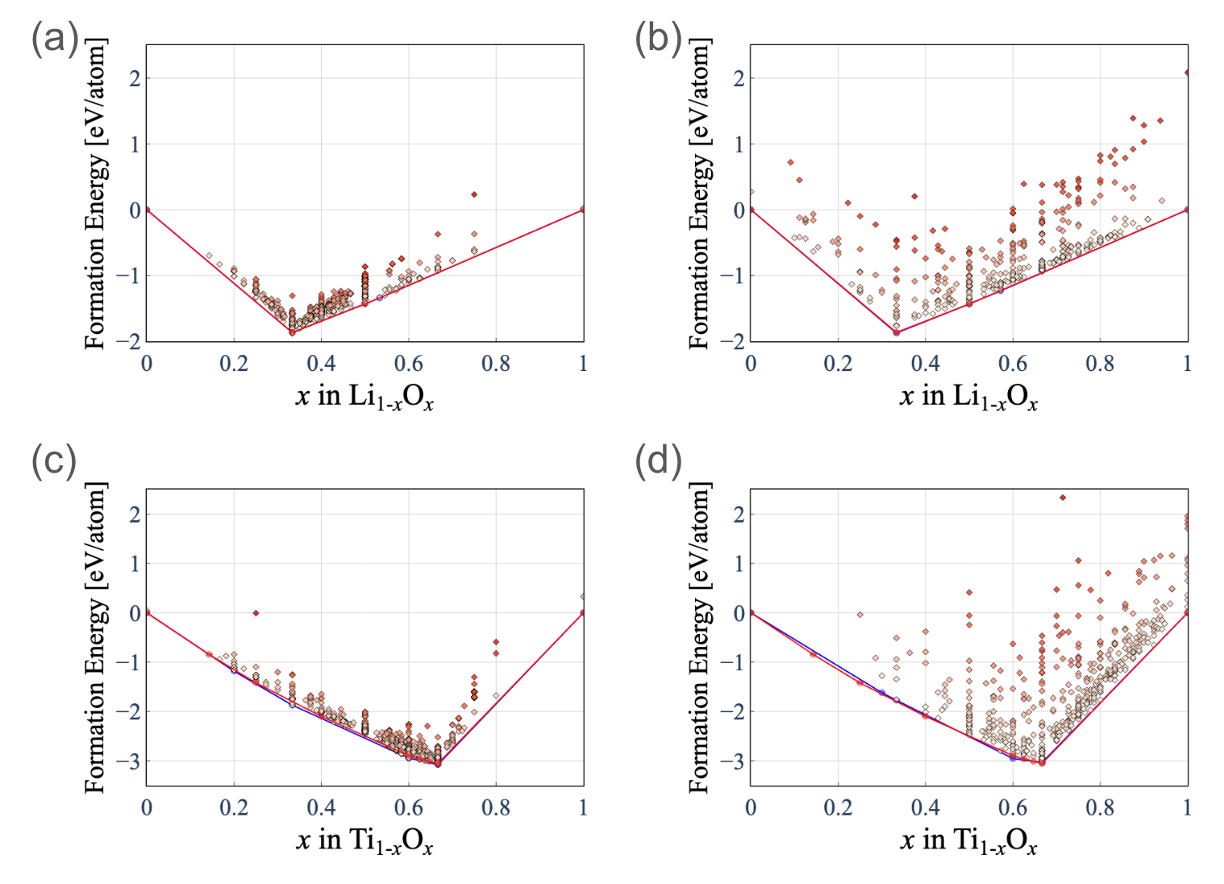}
\caption{Predicted phase diagrams of (a) Li–O by MatterGen, (b) Li–O by WyckoffDiff-Adaptor, (c) Ti–O by MatterGen, (d) Ti–O by WyckoffDiff-Adaptor,  Red and blue lines represent the convex hull of formation energy drawn by Materials project data and generated structures by the model, respectively.}
\label{fig:Li-O}
\end{figure*}

\begin{figure}
\centering
\includegraphics[width=0.95\linewidth]{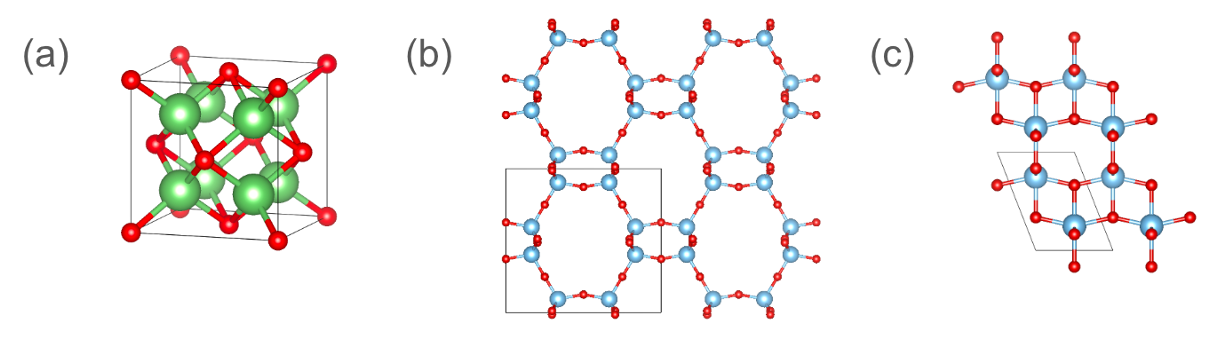}
\caption{(a) generated stable Li$_2$O structure
in  Li–O system shown in Figure 3 (b) , (b) generated stable TiO$_2$ structure shown in Figure 4 (b), and (c) stable TiO$_2$ structure in Materials project (mp-390).
}
\label{fig:structures_binary}
\end{figure}

Conditional generation of crystal structure not only controls the space groups but can also control the formation energy to explore stable phases for given element compositional space. 
Figure \ref{fig:Li-O} (a) and \ref{fig:Li-O} (b) show Li–O phase diagrams plotted using different specifications for the energy above hull = 0, with Li–O as the system under study and the top 20 most frequently occurring space groups from the dataset categorized by space group. Red lines represent stable structures present in the Materials Project \cite{Jain2013, Ong2008, Jain2011}, while blue lines indicate the hull formed by connecting stable structures generated using MatterGen. Each diamond point corresponds to a single system generated by the model.
Both of the Figures \ref{fig:Li-O} (a) and \ref{fig:Li-O} (b) show good agreement with the convex hull in the Materials Project data, clearly demonstrating that WyckoffDiff-Adaptor successfully explores a wide range of stable structures in the Li–O system. In addition, the generated structure by the WyckoffDiff-Adaptor reproduces the same configurations as the structure in the Materials Project database. Figure \ref{fig:structures_binary} (a) shows the structure of Li$_2$O crystal, which minimizes the formation energy in the phase diagram shown in Figure \ref{fig:Li-O} (b), which is similar to mp-1960.


Figure \ref{fig:Li-O} (c) and \ref{fig:Li-O} (d) show the phase diagram of the formation energy for Ti–O systems generated using the MatterGen, and the WyckoffDiff-Adaptor, respectively. In these cases, both of the models can succeed in exploring the convex hull of the stable structures. However, the stable structures are not identical to those in Materials project. Figure \ref{fig:structures_binary} (b) shows the generated stable structure of TiO$_2$. It is different from the well-known rutile structure shown in Figure \ref{fig:structures_binary} (c), which appears in the convex hull in the Materials project (mp-390). The task of exhaustive search of the convex hull is still difficult, even if the specific element and energetic conditions are given with the small number (1280 structures) of generated structures.

\begin{figure}
\centering
\includegraphics[width=0.8\linewidth]{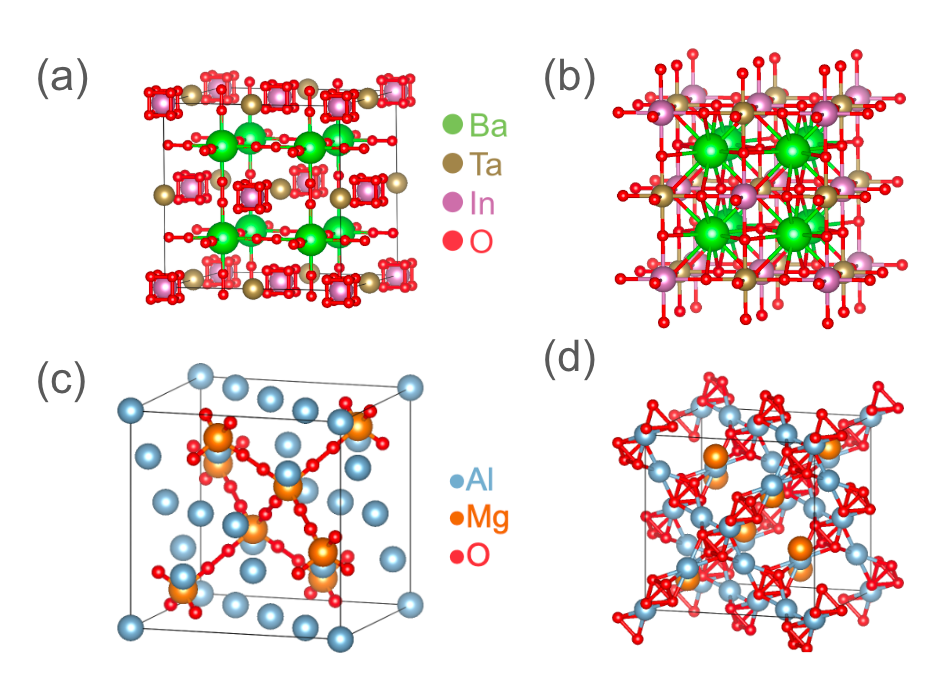}
\caption{(a)  an example in Ba-Ta-In-O system generated using WyckoffDiff-Adaptor, compared to the (b) structure in the same system in the Materials Project, and (c,d) structures generated by the post processing using PyXtal from the same protostructure (A2BC4$\_$cF56$\_$227$\_$c$\_$b$\_$e:Al-Mg-O) 
}
\label{fig:structures}
\end{figure}

Although the chemical element, spacegroup, and energetic conditions are successfully controlled by the conditional generation by the Adaptor module, WyckoffDiff does have some limitations. Figure \ref{fig:structures} (a) shows an example of these generated structures, under the space group conditioned with $Fm\overline{3}m$ (No. 225), with energy above the hull = 0, for the Ba-Ta-In-O system. The expected structure in the Materials project is shown in Figure \ref{fig:structures} (b), which is not in the training dataset (MP-20). The generated structures contain chemically unreasonable bond configurations with unnaturally neighboring O atoms around the Ta atoms. When these chemically implausible structures undergo structure optimization in MatterSim and subsequently phase diagram generation, they may exhibit higher energy values above the hull compared to those generated by MatterGen. Furthermore, in the heatmaps of space groups and crystal systems, the slightly increased occurrence of systems with lower symmetry  when specifying a space group likely stems from the  structure optimization in MatterSim with those inadequate initial generated structures, which significantly disrupted the structural symmetry after the optimization. The result suggests that the WyckoffDiff architecture may generate structures with insufficient assumption about the interaction among the atoms. 

Another significant drawback is that the method loses information about the internal parameters in generalized positions. For generalized positions, the coordinate system becomes ambiguous due to the values of these internal parameters. Nevertheless, WyckoffDiff generates only protostructures based on the Wyckoff Element Matrix, which do not contain any information about the internal parameters of generalized positions. Consequently, even in the original WyckoffDiff paper, the authors used PyXtal \cite{Fredericks2021} to semi-randomly determine internal parameters when generating crystal structures. This approach has the consequence that crystal structures containing important generalized positions tend to be difficult to generate. As an example, we demonstrate an attempt to generate the spinel-type structure MgAl$_2$O$_4$. Figures \ref{fig:structures} (c) and (d) show the two structures with the same protostructure (A2BC4$\_$cF56$\_$227$\_$c$\_$b$\_$e: Al-Mg-O), but with different oxygen site internal parameters generated randomly using PyXtal, resulting in distinct structural configurations. 

While the procedure of generating the protostructure can lead a problem of the arbitrary configurations with the protostructure, the computational cost for the generation is smaller, as it does not include the complete information on the coordinate and the lattice of the structure.

Table \ref{tab:performance} presents the timing results for generating CIF files using MatterGen and WyckoffDiff. As a benchmark, Table 1 shows performance data for 16 samples in the Li–O system, generated under controlled conditions using a V100 (16GB) GPU. MatterGen requires 366 seconds to generate cif files, whereas WyckoffDiff completes both prototype generation and the conversion from prototype to cif file in approximately 34 seconds and 26 seconds respectively, totaling about 60 seconds - roughly six times faster than MatterGen in cif file generation. This illustrates that while WyckoffDiff  presents challenges in handling highly flexible positions like generalized positions, the procedure can offer the advantage of rapid protostructure generation on the other hand. 

\begin{table}
\centering
\begin{tabular}{l|c c}
~ & MatterGen & WyckoffDiff \\\hline
Time [s] & 366 & 60(34(prototype) + 26(structure))
\end{tabular}
\caption{\label{tab:performance}Comparison of CIF file generation times between MatterGen and WyckoffDiff.}
\end{table}

\section{Conclusion}

In this study, we investigated the limitations of MatterGen and adopted WyckoffDiff as a model to address these limitations, enabling conditional generation and conducting experiments and analysis. The results revealed the following findings:
Using WyckoffDiff enables the generation of structures that preserve the specified space group, resulting in structures with higher symmetry compared to MatterGen.
WyckoffDiff can generate crystal structures significantly faster than MatterGen (approximately 6 times faster) while it cannot achieve the identical crystal structure with the degree of freedom of unconstrained Wyckoff positions.
In the Li–O system, we confirmed that it can produce stable structures comparable to those obtained using MP-20.
In addition, we observed cases where the model generates structures with high symmetry but chemically unfavorable bond configurations. Those structures may result in predictions of an unstable structure with symmetry lower than the input condition after the optimization of the structure.

\section*{Declaration of competing interests}
The authors declare that there is not any known competing financial interests or personal relationships that could have appeared to influence the work in this paper.

\section*{Acknowledgments}
The authors thank Akihide Hayashi and So Takamoto for fruitful discussions and helpful comments. 

\section*{Declaration of generative AI use}
During the preparation of this work the author(s) used the Gemini Pro (\url{https://gemini.google.com/}) and the Plamo translation (\url{https://translate.preferredai.jp/}) in order to revise English expressions. After using this tool/service, the author(s) reviewed and edited the content as needed and take(s) full responsibility for the content of the published article.

\section*{Data Availability}
The atomic structure data used for model training and comparative analysis are publicly and freely accessible. These structures were sourced from The Materials Project via its official API (\url{https://next-gen.materialsproject.org/api}) and specifically correspond to the contents of the MP-20 dataset \cite{xie2022}. All other data supporting the findings of this study are available from the corresponding author upon reasonable request.

\section*{Code Availability}
The implementation code for the diffusion model and the associated materials science application is available in a public repository at: \url{https://github.com/pfnet-research/wyckoffdiff_adaptor}. The results are based on existing foundational libraries: the MatterSim library, which is available as detailed in \cite{yang2024mattersim}, and the MatterGen framework is used for comparison of the model, which is available as detailed in \cite{MatterGen2025}.







\subsection{\bibliographystyle{elsarticle-num-names}}
\bibliography{ref}




\end{document}